\documentclass[prl,twocolumn,nopacs,nofootinbi]{revtex4-2}

\usepackage{graphicx}
\usepackage{dcolumn}

\usepackage{color}

\usepackage[dvips]{epsfig}
\usepackage{amssymb}

\usepackage{amsmath}

\usepackage{mathrsfs}
\usepackage{amsmath}
\usepackage{amssymb}
\usepackage{amsfonts}
\usepackage{latexsym}
\usepackage{color}

\usepackage{lipsum}

\newcommand{\bs}{\boldsymbol}

\newcommand{\markblue}[1]{{\leavevmode\color{black}{#1}}}

\begin{document}
\title{Self-learning mechanical circuits}

\author{Vishal P. Patil$^{1,2,\dagger}$}
\author{Ian Ho$^{1,\dagger}$}
\author{Manu Prakash$^{1,*}$}
\affiliation{$^1$Department of Bioengineering, Stanford University, Stanford, CA 94305, USA}
\affiliation{$^2$Department of Mathematics, University of California, San Diego, La Jolla, CA 92093, USA}
\affiliation{$^\dagger$ Equal contributions}
\address{* To whom correspondence should be addressed: manup@stanford.edu}

\date{\today}

\begin{abstract}

Adaptivity and learning are ubiquitous in living systems, from cytoplasm and biofilms to animal flocks. However, the problem of designing nonliving mechanical systems capable of self-learning behavior continues to present theoretical and experimental challenges. Here we introduce self-learning mechanical circuits: passive systems which take stochastic input forces from changing environments and constantly update their internal state in response, thus representing a mechanical unsupervised learning unit. Motivated by simplified models of  synapses, our circuits are composed of adaptive directed springs, which change stiffness in a directional manner. We provide an experimental realization of these elastic learning units and demonstrate how stability arises from adaptivity in certain network architectures. In particular, we theoretically demonstrate the ability of a network to autonomously uncover patterns in external inputs. Our results have applications towards the design of adaptive materials which can self-optimize in different environments.

\end{abstract}

\maketitle

\begin{figure}[h!]
	\centering	\includegraphics[width=\columnwidth]{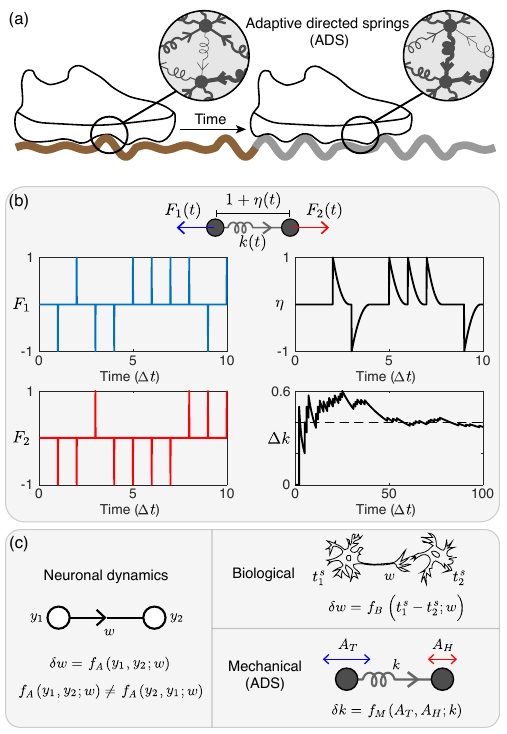}
	\caption{\textbf{Conceptualization of a self-learning mechanical circuit} 
    (a)~Toy example of a self-learning mechanical circuit embedded in a shoe walking over a textured surface (brown and grey curves). The network, made up of adaptive directed springs (ADSs), experiences input forces from the surface roughness, and responds by changing its internal spring stiffness pattern. When the surface patterning changes (grey curve), the network must learn a new stiffness pattern.
    (b)~Learning with an adaptive spring (equation~\ref{eq_toy_problem_learning}). The spring (inset) receives random forcing kicks $F_1$,$F_2$, (left column) at its endpoints, causing stretching, $\eta(t)$, (top right), and stiffness update, $\Delta k$, (bottom right, equation~\ref{eq_toy_problem_learning}). $\Delta k$ converges to the probability (dashed line) that the forcing kicks stretch the spring ($F_2>0, F_1<0$).
    (c)~Dynamics of neurons and synapses in artificial neural networks, as well as biological and mechanical networks. In its most generic form, neuron state can be a number $y_i \in \mathbb{R}$ (left), a most-recent spiking time $t^s_i$ (top) or an oscillation amplitude $A_i$ (bottom). In each case, typical learning rules ($f_{A}, f_{B}, f_{M}$) are non-symmetric functions of neuron state ($\markblue{y_i}, t^s_i, A_i$) and connectivity weights ($w, k$).
    }\label{fig1}
\end{figure}

\section{Introduction}


Mechanical principles give rise to adaptivity~\cite{stern2023learning} in a range of living~\cite{fleig2022emergence, kramar2021encoding,foster2019cytoskeletal, devany2021cell, rashid2023mechanomemory} and non-living systems~\cite{dillavou2022demonstration,vansaders2023informational,chen2021reprogrammable, zangeneh2021analogue, xia2022responsive, florijn2014programmable,wycoff2022desynchronous,dubvcek2024sensor}, from 
amorphous solids~\cite{stern2020continual,fiocco2014encoding,mukherji2019strength,keim2019memory, hagh2022transient} and deformable sheets~\cite{silverberg2014using,bhovad2021physical} to cells~\cite{pittman2022membrane, kim20224,lee2018multigenerational, nadell2015extracellular} and swarms~\cite{phonekeo2016ant, chakraborty2017swarm}. Using external computational optimization, tunable non-living materials can be engineered to have desired mechanical responses~\cite{lee2022mechanical, kwakernaak2023counting, reid2018auxetic, pashine2019directed, hexner2020periodic} to specific inputs~\cite{stern2020supervised, wright2022deep,altman2023experimental,de2026multimodal}. In contrast, living materials \markblue{can} adapt their mechanical properties autonomously, leading to functional consequences such as cellular contractility~\cite{squarci2023titin, foster2017connecting}, stress screening~\cite{mahuzier2018ependymal} and memory encoding~\cite{goult2021talin}. In particular, biological adaptive matter \markblue{can} self-learn by continually adapting itself to changing environments~\cite{bhattacharyya2022memory,ronellenfitsch2016global, grawer2015structural}, without the need for a central controller or an external reset. This raises the question of whether self-learning behavior can be embedded in the intrinsic dynamics of a driven elasto-mechanical system that is otherwise passive.

Here we construct self-learning mechanical circuits: passive elastic networks that are driven by external forces and continually adapt to mechanical information from this drive in an unsupervised fashion. Informally, a toy example of this behavior can be represented by an adaptive elastic network that receives stochastic input forces from its environment, through an interaction with a textured surface (Fig.~\ref{fig1}a, brown and grey curves). The network adapts to these inputs by changing its spring stiffnesses. \markblue{To achieve the goal of learning information} about the environmental forcing inputs, these spring stiffnesses should encode parameters of the stochastic input forces. The learned, internal state of the network is then its spring stiffness pattern. To capture the continual adaptivity of biological systems, this learned state must also change when the environment changes (Fig.~\ref{fig1}a, brown and grey curves).

To realize this form of self-learning dynamics we construct adaptive springs, which are the fundamental unit of our mechanical circuits. Motivated by a theoretical example of learning using elastic springs (Fig.~\ref{fig1}b), we engineer our adaptive springs to be directed, with stiffness update depending on the force difference across the spring. Further, the resulting adaptive directed springs (ADSs) are battery-free, and only adapt in response to external forcing. These elastic units can be thought of as mechanical objects which capture certain aspects of neuronal and synaptic dynamics. Crucially, this learning behavior is unsupervised, arising from the intrinsic adaptive dynamics of the mechanical circuit. Below we construct these springs and demonstrate theoretically their application to unsupervised learning problems.

\section{Learning in elastic networks}

To identify the relevant forms of adaptivity, we first formalize a class of self-learning problems in elastic networks. We treat our mechanical circuits as dynamical systems which experience stochastic forces and learn features of the underlying force distribution. The state of the dynamical system can be split into a set of physical degrees of freedom, $\mathbf{X}$, and internal or learning degrees of freedom, $\mathbf{K}$. The mechanical networks studied here are composed of masses and springs, with $\mathbf{X}$ denoting the positions and possibly the velocities of all the masses, and $\mathbf{K}$ denoting the spring stiffnesses. The dynamical system interacts with its environment through a time-dependent stochastic forcing term $\mathbf{F}^\text{env}(t)$ which directly affects the physical degrees of freedom, $\mathbf{X}$, but not the learning degrees of freedom, $\mathbf{K}$. We therefore write
\begin{equation}
\begin{aligned} \label{eq_abstract_learning}
    \dot{\mathbf{X}} &= f_X(t, \mathbf{X}, \mathbf{K} \,;\, \mathbf{F}^{\text{env}}) \\
    \dot{\mathbf{K}} &= f_K(t, \mathbf{X}, \mathbf{K})
\end{aligned}
\end{equation}
We further focus on passive systems, with the property that $\dot{\mathbf{X}}, \dot{\mathbf{K}} \rightarrow 0$ as $t\rightarrow \infty$ whenever $\mathbf{F}^{\text{env}} = 0$. In our case, $f_X$ represents the elastic network dynamics, and $f_K$ is a learning rule by which the spring stiffnesses of the network update.

To formulate a learning problem, we assume that the distribution of $F^{\text{env}}(t)$ depends on a \markblue{parameter} $\bs{\Phi}$. The fundamental learning objective is to estimate features $\mathcal{P}(\bs{\Phi})$ of $\bs{\Phi}$, such as the force covariance matrix $\langle \mathbf{F}^{\text{env}} \otimes  \mathbf{F}^{\text{env}} \rangle$.  To do this in an unsupervised fashion, the estimate must be generated using only the dynamics of the network. Unsupervised learning therefore requires constructing an estimator $\hat{P}$ for $\mathcal{P}(\bs{\Phi})$ which depends only on $\mathbf{K}$ and $\mathbf{X}$ at some time $T$, so $\hat{P} = \hat{P}\left(\mathbf{K}(T), \mathbf{X}(T) \right)$. In general, the dynamics of equation~\eqref{eq_abstract_learning} will not be able to learn complete information about $\bs{\Phi}$ for fixed functions $f_X,f_K$. Choosing functions $f_X, f_K$ in order to learn particular features of $\bs{\Phi}$ is akin to choosing an unsupervised learning algorithm~\cite{hastie2009elements}. In the context of elastic networks, $f_X$ can be changed by modifying the connectivity of the network and $f_K$ is set by the stiffness update dynamics of the springs. In addition to being unsupervised, the learning dynamics are powered by the signal $\mathbf{F}^{\text{env}}(t)$, since $\mathbf{K}, \mathbf{X}$ are driven by $\mathbf{F}^{\text{env}}(t)$. The system~\eqref{eq_abstract_learning} therefore describes learning without a battery.

The distinction between a classical elastic network and an adaptive, self-learning network can be illustrated through a simple example of the unsupervised learning problem described above. Consider a single overdamped spring in 1D connecting two nodes \markblue{with positions $x_1(t), x_2(t)$,} subject to environmental forces. To formulate a learning problem, we set the environmental forcing term on each node to be a sequence of random forcing kicks at $\Delta t$ time intervals  (Fig.~\ref{fig1}b). The profile of a kick at time $t_0$ is given by a regularized delta function, \markblue{$F_\delta(t)$}, which is $1/\delta$ for $0<t<\delta$ and $0$ elsewhere, with $\delta/\Delta t \ll 1$. Working in dimensionless units, the force on node $i$ is then 
\markblue{
\begin{align}\label{eq_env_force_toy}
F^{\text{env}}_i(t) = \sum_j Y_i^j F_\delta(t - j \Delta t)
\end{align}
}
where $Y_1^1, Y_1^2,Y_1^3...$ and $Y_2^1, Y_2^2,Y_2^3...$ are the strengths of the kicks on the two nodes and each $Y_i^j \in \{-1,+1\}$. Both of the sequences $\{Y_1^j\}$ and $\{Y_2^j\}$ are individually i.i.d., but there are correlations between them. In particular, let the joint probability that $(Y_1^j, Y_2^j)$ takes the values $(-1,-1), (-1,1), (1,-1), (1,1)$ be $p_{11}, p_{12}, p_{21}, p_{22}$ respectively, where $\sum p_{ij} = 1$. The input forces $\{F^{\text{env}}_i(t) \}$ therefore depend on a \markblue{parameter} $\bs{\Phi} = (p_{11},p_{12},p_{22}) $. To ensure that $(Y_1^j, Y_2^j)$ are not independent, we set
\begin{align*}
    \mathbb{P}\{(Y_1^j, Y_2^j)=(-1,-1)\} &\neq \mathbb{P}\{Y_1^j = -1\} \, \mathbb{P}\{Y_2^j = -1\}
\end{align*}
or equivalently, $p_{11} \neq (p_{11}+p_{12})(p_{11}+p_{21})$.
In dimensionless units, the overdamped dynamics of the spring can then be written in terms of the spring extension \markblue{$\eta = x_2 - x_1$} and center of mass \markblue{$x_{\mathrm{cm}} = (x_1 + x_2)/2$
\begin{equation}
\begin{aligned}
\label{eq_1spring_toy}
    \dot{\eta} &= -2k(t)U'(\eta) + \sum_j (Y_2^j - Y_1^j)F_\delta (t - j\Delta t)  \\ 
    \dot{x}_{\mathrm{cm}} & = \frac{1}{2}\sum_j (Y_2^j + Y_1^j)F_\delta (t - j\Delta t) 
\end{aligned}
\end{equation}
}
where $k(t)$ is the spring stiffness, and $kU(\eta)$ is the potential energy of the spring. In the notation of equation~\eqref{eq_abstract_learning}, the physical degrees of freedom are \markblue{$\mathbf{X} = (\eta, x_{\mathrm{cm}})$} and the learning degrees of freedom are $\mathbf{K} = (k)$.

A non-adaptive spring experiencing the above driving protocol \markblue{can learn partial information}, but is unable to learn complete information about its environmental forces. This can be seen from equation \eqref{eq_1spring_toy} for a constant-stiffness spring with $k(t) \equiv k_0$. Integrating the center-of-mass equation up to time $T$ gives
\markblue{
\begin{align*}
    x_{\mathrm{cm}}(T) = x_{\mathrm{cm}}(0) + \frac{1}{2}\sum_{j=1}^{N} (Y_2^j + Y_1^j)
\end{align*}
}
where $N = \lfloor T/\Delta t \rfloor$, which captures the overall drift of the spring. Since $\langle Y_2^j + Y_1^j\rangle = 2(p_{22} - p_{11})$, the center-of-mass equation therefore gives \markblue{$\langle x_{\mathrm{cm}}(T) - x_{\mathrm{cm}}(0) \rangle = N(p_{22} - p_{11})$. However, the distribution of $x_{\mathrm{cm}}(T)$} does not depend on $p_{12}, p_{21}$, so cannot give any information about these joint probabilities. The fact that $\eta$ also fails to learn these quantities follows from its relaxation dynamics. Provided $\Delta t$ is large enough, $\eta$ approximately returns to $0$ between forcing kicks. For a linear spring ($U(\eta) \propto \eta^2$) in the presence of friction-type nonlinearities, it is reasonable to expect $\eta \approx 0$ in between kicks. In Fig.~\ref{fig1}b, we take a nonlinear spring energy, $U(\eta) = 2|\eta|^{3/2}/3$, which has the property that $\eta$ hits $0$ in finite time, and $\eta(j\Delta t - \epsilon) = 0 $, for small $\epsilon > 0$. If the system is stopped just before the $(N+1)$'th kick, at time $T = (N+1)\Delta t - \epsilon$, then $\eta = 0$ and therefore contains no information about the forcing parameters $p_{ij}$.  This non-adaptive system therefore cannot learn $\bs{\Phi}$.

By allowing the spring stiffness to update as a function of the node \markblue{motion} (equation~\ref{eq_abstract_learning}), the system can also learn the parameter $p_{12}$, \markblue{in addition to $p_{22} - p_{11}$}. This sets the probability of $\{(Y_1^j, Y_2^j) = (-1,+1)\}$, corresponding to a stretch of the spring. To compute $p_{12}$, consider a spring which increases in stiffness whenever it is stretched, and has unit stiffness at rest. The system in \eqref{eq_1spring_toy} can be augmented with such a learning rule
\begin{equation}
\begin{aligned}\label{eq_toy_problem_learning}
    \dot{k} = -\frac{1}{t}(k-1) + \frac{\Delta t}{t\delta} \Theta\left(\dot{\eta} - 1\right)
\end{aligned}
\end{equation}
where $\Theta$ is the Heaviside step function. $k$ therefore tends to increase when $\dot{\eta}>1$, which occurs when $(Y_2^j - Y_1^j) > 0$.  Integrating the $k$ equation with $k(0) = 1$ gives
\begin{align*}
    k(t) = 1 + \frac{\Delta t}{t\delta} \int^t \Theta(\dot{\eta}-1) dt'
\end{align*}
Since $k \geq 1$, the magnitude of the restoring force on $\eta$ is at least $2U'(\eta)$ (equation \ref{eq_1spring_toy}). Assuming that $\eta \approx 0$ between forcing kicks, we have that $\dot{\eta}>1$ for time $\delta$ only when $(Y_2^j - Y_1^j) = 2$, which occurs with probability $p_{12}$. The expected stiffness just before the $(N+1)$'th kick, at time $T = (N+1)\Delta t - \epsilon$, then satisfies
\markblue{\begin{align}
    \langle k(T) \rangle = 1+ \frac{\Delta t  }{ T\delta }  N\delta \, p_{12} = 1+ \frac{N\Delta t}{T}p_{12}
    \label{eq_k_learns_p12}
\end{align}
Furthermore, as $N \rightarrow \infty$}, $\Delta k  = k(T) - k(0) \rightarrow p_{12}$ (Fig.~\ref{fig1}b). By encoding the joint probability, the stiffness therefore learns a non-trivial parameter of the joint distribution of node forces, thus performing an unsupervised learning task.

\markblue{The above example demonstrates that stiffness updates can encode statistical information about driving forces. To motivate our physical realization of learning springs, we emphasize three key features of this example. The first is the adaptive stiffness itself, which enables the spring to learn more parameters of the input force (equation \ref{eq_k_learns_p12}). Secondly, the system utilizes threshold nonlinearities, through the Heaviside step function $\Theta$ in the learning rule (equation~\ref{eq_toy_problem_learning}). The third feature is that the learning rule is directional, meaning that $\dot{k}$ depends non-symmetrically on each node, due to the term $\Theta(\dot{\eta}-1) = \Theta(\dot{x}_2 - \dot{x}_1 - 1)$ in equation~\eqref{eq_toy_problem_learning}. Directionality is also crucial in simplified descriptions of biological and artificial neurons~\cite{caporale2008spike, legenstein2008learning, mandal2024learning} (Fig.~\ref{fig1}c).

Here we specialize to the case where a network is driven periodically, and focus on constructing the simplest ADS which can compare the motion of its two endpoints. In this case the state of each node is its oscillation amplitude along the direction of the spring $\hat{\bs{b}}$ (Fig.~\ref{fig1}c). To emphasize directionality, we assign our adaptive springs a tail node and head node, with displacements $\bs{u}_T, \bs{u}_H$ from rest, respectively. The oscillation amplitudes on each node are then  $A_T = \langle |\bs{u}_T \cdot \hat{\bs{b}}| \rangle $  and $A_H = \langle |\bs{u}_H \cdot \hat{\bs{b}}| \rangle $. To compare endpoint motion, the stiffness $k$ of the ADS is set to continually update if $A_T > A_H$, but remain constant if $A_T\leq A_H$
\begin{subequations}\label{ads_axioms_2}
\begin{alignat}{2}
    &\dot{k} \not\equiv 0 \quad\text{if}\; A_T > A_H \quad \text{Directional update} \\ 
    &\dot{k} \equiv 0 \quad\text{if}\; A_T \leq A_H \quad \text{Self-regulation}
\end{alignat}
\end{subequations}
The form of adaptivity is loosely analogous to spike-timing-dependent plasticity \cite{caporale2008spike, legenstein2008learning,mandal2024learning}, where synaptic weight changes depend on the relative activity of pre- and post-synaptic neurons (Fig.~\ref{fig1}c). Taken together, these features balance adaptivity and stability. We next demonstrate how to construct a physical realization of this ADS.
}

\section{Physical realization of adaptive directed springs}

\begin{figure*}[t]
	\centering	\includegraphics[width=\textwidth]{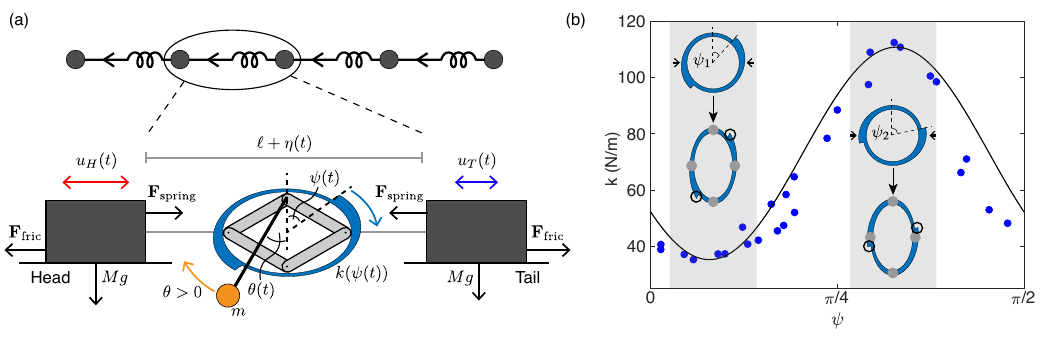}
	\caption{\textbf{Schematic of an adaptive directed spring (ADS).}  
    (a)~Free body diagram of an ADS between two nodes  (black rectangles). A pendulum (orange) and an elastic ring (blue) are mounted on a four-bar linkage (light grey). The ring is ratcheted and can only rotate in one direction, $\dot{\psi} \geq 0$ (blue arrow). The spring stiffness, $k$, is a function of $\psi$. The positive direction of the pendulum \markblue{angle} (orange arrow) points from tail to head by definition. The spring stiffness, $k$, depends on the node oscillations, $u_T,u_H$, through the pendulum angle, $\theta$ and the spring extension $\eta$ (the spring rest length is denoted by $\ell$). $k$ updates only when $\dot{\eta}, \dot{\theta} > 0$. Friction acts as a threshold nonlinearity on the spring.
    ~(b) Experimentally measured spring stiffness (blue circles) as a function of $\psi$ with lowest Fourier mode approximation (equation~\ref{eq_spring_stiffness}) overlaid (solid line). The force required to compress the elastic ring depends on ring angle $\psi$. The distance between the thickest portions of the deformed ring (black circles, inset) and the portions under greatest strain (grey circles, inset) determines spring stiffness: stiffness is large when this distance is small.}  
	\label{fig2}
\end{figure*}
We outline a physical instantiation of the ADS conceptualized above which relies upon three mechanical ingredients: substrate friction which produces a threshold nonlinearity, an elastic ring of non-uniform thickness which provides variable stiffness~\cite{umedachi2006development}, and a coupled pendulum-linkage system to enable adaptivity in a directional manner (Fig.~\ref{fig2}). Taken together, the oscillation of the passive pendulum-linkage system, when driven by the environment, causes the elastic ring to rotate thus changing the stiffness of the ADS. \markblue{These components allow this realization of an ADS to exhibit
threshold activation, adaptive stiffness, and the directional
update in equation~\eqref{ads_axioms_2}.}


We begin by demonstrating \markblue{threshold activation and adaptive stiffness.} A single ADS consists of nodes of mass $M$ and displacements $u_i, u_j$ on a substrate, which interact elastically via an elastic ring with varying thickness (Fig.~\ref{fig2}a, blue ring). The elasticity of the ring gives rise to the stiffness of the spring, $k$, which changes as the ring rotates. Coulomb friction~\cite{shaw1986dynamic,zhou2014energy} on each node functions as a threshold nonlinearity (Fig.~\ref{fig2}a). For example, the equation of motion \markblue{for the node displacement $u_j$ is}
\begin{align}\label{eq_friction}
    M\ddot{u}_j = -b\dot{u}_j - k(u_j - u_i)+F^f
\end{align}
where $-b\dot{u}_j - k(u_j - u_i) = F^0 $ is the \markblue{total non-friction force} on node $j$, arising from spring elasticity and a viscous drag $-b\dot{u}_j$, and $F^f$ is the friction force. Following references~\cite{shaw1986dynamic,zhou2014energy}, the form of Coulomb friction chosen here is 
\begin{align}\label{eq_friction_2}
F^f = \begin{cases} 
-\mu M g \, \text{sgn}(\dot{u}_j) \quad \text{if } |\dot{u}_j| > 0 \\
-\min(\mu M g, |F^0|)\, \text{sgn}(F^0) \quad \text{if } |\dot{u}_j| = 0 \\
\end{cases}
\end{align}
where $g$ is gravitational acceleration. The total force, $F = F^0+F^f$, then behaves as a threshold, with $F = 0$ when $|\dot{u}_j|=0$ unless $F_0$ is sufficiently large. The activation nonlinearity is therefore captured by $F$.

The rotation angle, $\psi$, of the elastic ring determines the spring stiffness $k = k(\psi)$, and is responsible for regulating the transmission of forces throughout the network. When compressed, the absolute change in curvature along the ring, $|\delta \kappa|$, has four maxima (Fig.~\ref{fig2}b). The lowest Fourier mode approximation for the stiffness function is therefore
\begin{align}\label{eq_spring_stiffness}
	k(\psi) = k^0 + \frac{k_{\max} - k_{\min}}{2} \cos \left( 4\psi + \alpha_0 \right)
\end{align}
where $k^0 = ({k_{\max} + k_{\min}})/{2}$ and $\alpha_0$ is a constant which depends on the choice for $\psi = 0$ on the ring. Experimental measurements of spring stiffness as a function of $\psi$ confirm that $k(\psi)$ has period $\pi/2$ (Fig.~\ref{fig2}b, SI). The thickness profile of the elastic ring used here (Fig.~\ref{fig2}a) is chosen to aid experimental tracking of the rotation angle $\psi$ (SI). However, other non-uniform thickness profiles will also have $k-\psi$ relationships of the form given in \eqref{eq_spring_stiffness}. 
Together, friction and the elastic ring \markblue{provide threshold
activation and adaptive stiffness.}

When driven by external oscillatory forces, the passive ADS mechanism exhibits a directional stiffness update, which arises from phase-coupling in a pendulum-linkage system. To explain the stiffness update rule, we first describe these phase-coupling dynamics in terms of node oscillations. Here, the pendulum \markblue{angle} is denoted by $\theta$, and the spring extension by $\eta$ (Fig.~\ref{fig2}a). Choosing a positive direction for the pendulum angle $\theta$ gives the spring a direction (Fig.~\ref{fig2}a). The head and tail nodes are chosen so that the tail-to-head vector points along this axis. In terms of the head and tail node displacements, $u_H, u_T$, the spring extension is then $\eta = u_H - u_T$. The equation of motion of the pendulum is
\begin{align}\label{eq_oscillator_full}
    \ddot{\theta} +b_p\dot{\theta} = -\omega_p^2\theta - \frac{1}{2\ell_p}\left(\ddot{u}_T + \ddot{u}_H\right)
\end{align}
where $b_p$ is a damping coefficient, $\ell_p$ is the pendulum length and $\omega_p = \sqrt{g/2\ell_p}$ is the natural frequency. The $-(\ddot{u}_T + \ddot{u}_H)$ term captures the additional torques due to node accelerations, $\ddot{u}_{T,H}$. When damping is small and the typical node accelerations are much larger than $g$, we can neglect the gravitational torque, giving the following dynamics
\begin{align}\label{eq_oscillator}
    \ddot{\eta} = -\ddot{u}_T + \ddot{u}_H \;,\quad 2\ell_p\ddot{\theta} = -\ddot{u}_T - \ddot{u}_H
\end{align}
for $\eta$ and $\theta$.

Directional stiffness update requires an update rule which depends on odd powers of $\eta$ and its derivatives, as in equation~\eqref{eq_toy_problem_learning}. The ADS can satisfy this property by updating only when $\dot{\theta}$ and $\dot{\eta}$ are both positive. Physically, this corresponds to stiffness update only when the oscillation amplitude of the tail node exceeds the head node. To see this, suppose each node is oscillating with frequency $\omega$, and amplitude $A_{T,H}$, so $u_{T,H} = A_{T,H}\cos\omega t$. This gives $\eta \sim (-A_T + A_H)\cos\omega t$ and $\theta \sim (-A_T - A_H)\cos\omega t$. Thus $\theta, \eta$ oscillate in-phase if $A_T > A_H$ and in antiphase otherwise. Thus $\dot{\theta},\dot{\eta}$ are simultaneously positive only when $A_T > A_H$, corresponding to in-phase oscillations. We next show how to engineer the ADS so that its stiffness updates only when $\dot{\theta},\dot{\eta} >0$.

\begin{figure}[t]
	\centering	\includegraphics[width=\columnwidth]{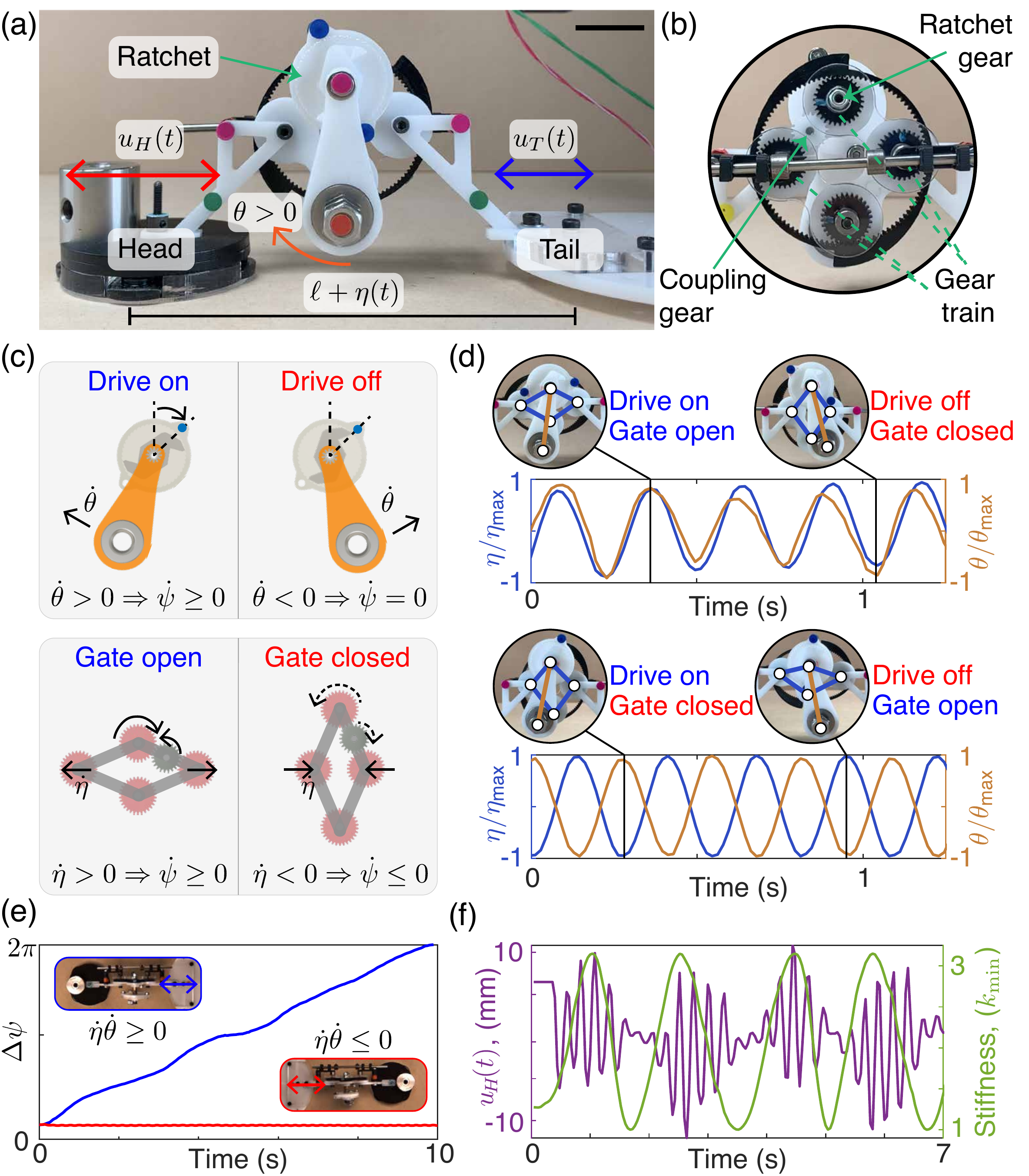}
	\caption{\textbf{Physical construction of an adaptive directed spring (ADS).}  
    (a)~Implementation of an ADS. The $\theta>0$ direction points from tail to head. A ratchet is coupled to the pendulum. Scale bar~$2\,$cm. (b) Back: The inner rim of the elastic ring is patterned with gear teeth and rests on a gear train. The gear train includes a ratchet gear, which is directly coupled to the pendulum. Rotation of the ratchet gear updates $\psi$.
    (c)~The ring angle $\psi$ responds differently to $\theta$ and $\eta$. When $\dot{\theta}>0$, the pendulum (orange) engages the ratchet (drive on, $\dot{\psi}\geq 0$), otherwise $\dot{\psi} = 0$ (top row). The gear mechanism (bottom row) functions as a gating mechanism. When the linkage is stretched (compressed), the gear torques and constraints only permit the ratchet-gear to rotate clockwise (counterclockwise), forcing $\dot{\psi} \geq 0$ ($\dot{\psi} \leq 0$), and leading to the gate open (closed) state. Only the drive on, gate open state leads to spring update, $\dot{\psi} > 0$ (SI). 
    (d)~Oscillating a physical ADS from the tail (top row) produces in-phase $\eta,\theta$ dynamics, whereas input head oscillations (bottom row) cause $\eta,\theta$ to be in antiphase. (e)~In-phase motion hits the drive on, gate open state ($\dot{\theta},\dot{\eta}>0$), leading to $\dot{\psi}>0$ (blue curve). Antiphase motion leads to $\dot{\psi}=0$ (red curve).
    (f)~Elastic response of a physical ADS under input tail oscillations displays the hallmarks of adaptivity and thresholding (movie~S1). The changing stiffness (green curve) causes the output amplitude (purple curve) to increase and decrease.
    } 
	\label{fig3}
\end{figure}

\begin{figure*}[t]
	\centering	\includegraphics[width=\textwidth]{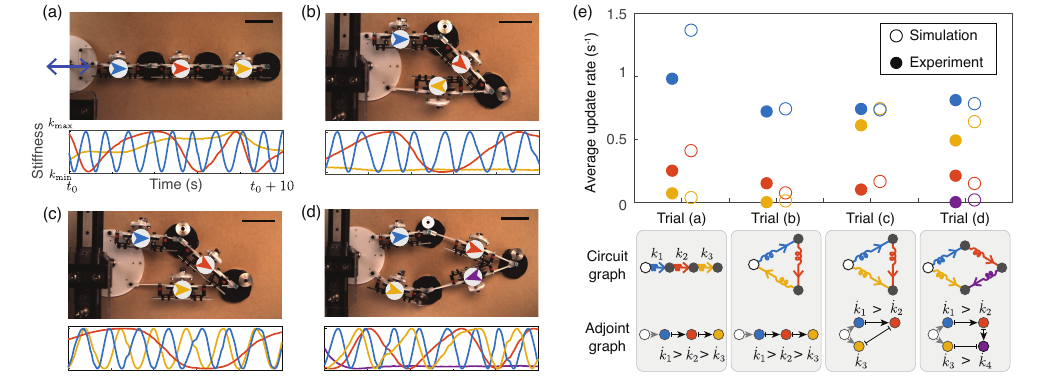}
	\caption{\textbf{Dynamics of self-learning mechanical circuits.} 
    (a-d)~Physical mechanical circuits (top) of 3 and 4 ADSs produce a variety of distinct learning dynamics (movie~S3), as revealed by their spring stiffness time series (bottom). The input node (large white disk) oscillates at constant amplitude ($\approx 1\,$cm) and frequency ($\approx 3\,$Hz); the remainder of the network is purely mechanical. Arrows point from the tail to the head of each ADS. The springs are free to rotate around their nodes except for circuit (a) where the spring angles are mechanically fixed. Changing the direction of a single ADS (b,c; yellow arrows) dramatically alters the learning dynamics (second row). Scale bars~$6\,$cm. (e)~Equations~\eqref{eq_elastic_network_force}–\eqref{eq_learning_network} agree quantitatively with experimentally observed update rates for circuits in (a-d) (full simulated stiffness time series provided in SI). Bottom row: features of the learning dynamics can be inferred from graph representations of mechanical circuits (input forcing node colored white). Adjoint graphs represent ADSs as nodes (colored circles) receiving input forces (white circle, short grey arrows). Arrowhead style denotes promoters (sharp arrowhead) or inhibitors (flat arrowhead). In column 1, the blue ADS meets the red ADS at its tail (promoter), but the red ADS meets the blue ADS at its head (inhibitor). Edges further downstream of the input forces (white circles) update more slowly and thus store information about the long timescale variations of the input signal.
 } 
	\label{fig4}
\end{figure*}

In order to realize directional stiffness update in an ADS, we design the system to exhibit different states depending on the signs of $\dot{\theta}$ and $\dot{\eta}$. This is achieved by coupling a centripetal ratchet to the pendulum and a gear train to the linkage, whose state is set by $\eta$ (Fig.~\ref{fig3}a,b). The ratchet and gear train are connected to each other via a `ratchet-gear' (Fig.~\ref{fig3}b), whose net rotation directly drives the unidirectional update of the elastic ring ($\psi$). When $\dot{\theta}>0$, the pendulum drives the ratchet and in turn rotates the ratchet-gear, increasing $\psi$ (Fig.~\ref{fig3}c, top left). However, if $\dot{\theta}<0$, the unengaged ratchet remains stationary and $\dot{\psi}=0$ (Fig.~\ref{fig3}c, top right). In contrast, the stretching of the linkage does not produce enough torque in the gear train to directly drive the ratchet-gear, due to inertia (SI). Stretching instead enforces a gear rotation pattern which either permits ($\dot{\eta}>0$) or prevents rotation ($\dot{\eta}<0$) of the ratchet-gear (Fig.~\ref{fig3}c bottom row) due to the existence of a symmetry-breaking `coupling gear' in the gear train (Fig.~\ref{fig3}b). When the linkage and spring stretch, the ratchet-gear can either rotate clockwise or remain stationary ($\dot{\psi}\geq 0$), whereas when the linkage and spring compress, the ratchet-gear can either rotate counterclockwise or remain stationary ($\dot{\psi} \leq 0$). In this way, $\eta$ acts as a gating mechanism for the spring update (Fig.~\ref{fig3}c, bottom row). Therefore, $\dot{\psi} > 0$ requires the ADS to be in both the ``drive on'' ($\dot{\theta} > 0$) and ``gate open'' ($\dot{\eta}>0$) states (Fig.~\ref{fig3}b), and the ADS learning rule takes the form
\begin{align}\label{eq_learning_rule}
    \dot{k} = \gamma k'(\psi) \, \text{Relu}(\dot{\theta}-\tau_0) \, \Theta(\dot{\eta}-\tau_1)
\end{align}
where the learning rate, $\gamma$, is set by the gear ratios involved in the design, $\Theta$ is the Heaviside step function, and $\tau_0$ and $\tau_1$ are additional thresholds empirically included to capture the effects of internal friction. The directionality of the ADS can be modified through rearrangements of the gear-linkage system (SI).

We experimentally validate that an ADS exhibits both adaptive and directional behavior (Fig.~\ref{fig3}d,e,f). Using a linear translation stage, the tail or head node of an ADS can be directly oscillated, while keeping the other node free. Measuring $\theta$, $\eta$ and $\Delta \psi$ when the tail node is driven (i.e. $\langle |u_T| \rangle > \langle |u_H| \rangle$) confirms that in-phase oscillations of $\theta$ and $\eta$ (Fig.~\ref{fig3}d, top row) lead to $\Delta \psi > 0$ (Fig.~\ref{fig3}e, blue) thus changing the spring stiffness (Fig.~\ref{fig3}f). On the other hand, when the head node is driven (i.e. $\langle |u_T| \rangle < \langle |u_H| \rangle$), antiphase $\eta, \theta$ oscillations are observed (Fig.~\ref{fig3}d, bottom row) which fix the spring stiffness (Fig.~\ref{fig3}e, red). Here, \markblue{the pendulum} is set in motion from rest and reaches its steady oscillation rapidly $t\leq 0.5\;\text{s}$ (movie~S1,2), and thus transient motion in the pendulum does not significantly affect learning dynamics. Furthermore, provided the oscillation frequency is far from resonance for either $\theta$ or $\eta$, this phase behavior is robust to perturbations (movie~S2). Update in an ADS happens by harvesting external vibrations, so no internal power source is needed. This construction thus enables adaptive behavior in an otherwise passive spring.

\section{Self-learning mechanical circuits}

Connecting multiple ADSs produces mechanical circuits with rich landscapes of limit cycles and fixed points, reminiscent of networks of neurons. When excited at a node, forces propagate through a mechanical circuit, causing the stiffness of certain ADSs to update (Fig.~\ref{fig4}a-d, movie~S3). The learning dynamics can be described using the friction-oscillator model from equations \eqref{eq_friction}$-$\eqref{eq_learning_rule}, over an elastic network. To write down these equations, consider a general elastic network with nodes at $\mathbf{r}_i + \bs{u}_i \in \mathbb{R}^2$, where $i \in V = \{1,2,...,N\}$, $\bs{r}_i$ is the rest position of node $i$ and $\bs{u}_i$ is its displacement. The network is a directed graph $G=(V,E)$, where the edge set $E$ consists of ordered pairs of nodes, with $e=(i,j) \in E$ if there is an ADS with tail end at $i$ and head end at $j$. The edge vectors are denoted $\mathbf{b}_{ij} =\mathbf{r}_j  -\mathbf{r}_i$ and $\hat{\mathbf{b}}_{ij} = \mathbf{b}_{ij} / |\mathbf{b}_{ij}|$. In general, the nodes can experience elastic forces, $\mathbf{F}_i^{\text{elast}}$ and external forces, $\mathbf{F}_i^{\text{ext}}$ as well as a viscous drag $-b\dot{\mathbf{u}}_i$ and nonlinear friction. The total non-friction force on node $i$ is $\mathbf{F}^0_i = -b\dot{\mathbf{u}}_i + \mathbf{F}_i^{\text{elast}} + \mathbf{F}_i^{\text{ext}}$. \markblue{To write down the elastic force, we first define the set of neighbors of node $i$ as $N_b(i) = \{j\in V:(i,j)\in E\ \text{or}\ (j,i)\in E\}$. For $j\in N_b(i)$, let $k_{ij}$ denote the instantaneous physical stiffness of the ADS connecting nodes $i$ and $j$. The stiffness is independent of head--tail orientation, so $k_{ij} = k_{ji}$. For relative node displacements that are small compared to the spring length, $\lVert\bs{u}_j-\bs{u}_i\rVert\ll\lVert\bs{b}_{ij}\rVert$, the leading-order elastic force on node $i$ is \cite{lubensky2015phonons}
\begin{equation}
\bs{F}^{\mathrm{elast}}_i
=
-\sum_{j\in N_b(i)}
k_{ij}
\left[
\hat{\bs{b}}_{ij}\cdot
(\bs{u}_i-\bs{u}_j)
\right]
\hat{\bs{b}}_{ij}.
\label{eq_elastic_network_force}
\end{equation}
}


The friction force, $\mathbf{F}^f\left(\mathbf{u}_i, \mathbf{F}^0_i \,;\,\mu Mg \right)$ on a node follows from a vector version of equation~\eqref{eq_friction_2}
\begin{align} \label{eq_friction_network}
    \mathbf{F}^f  = \begin{cases} -\mu M g \,\dfrac{\dot{\boldsymbol{u}}_i}{|\dot{\boldsymbol{u}}_i|} \quad \text{if } |\dot{\boldsymbol{u}}_i| > 0 \\
    -\min \left( \mu M g, |\mathbf{F}^0_i| \right) \, \dfrac{\mathbf{F}^0_i}{|\mathbf{F}^0_i|} \quad \text{if } |\dot{\boldsymbol{u}}_i| = 0
    \end{cases}
\end{align}
The elastic equations for an ADS network with friction are thus
\begin{align}\label{eq_elastic_network}
\begin{split}
    M\ddot{\mathbf{u}}_i &= \mathbf{F}^0_i+ \mathbf{F}^f(\dot{\mathbf{u}}_i, \mathbf{F}^0_i \,;\; \mu M g) \\
    \mathbf{F}^0_i &= -b\dot{\mathbf{u}}_i + \mathbf{F}_i^{\text{elast}} + \mathbf{F}_i^{\text{ext}}
\end{split}
\end{align}
Finally, the learning dynamics for the stiffnesses, $k_{ij}$, follow from the stretch, $\eta_{ij}$, pendulum angle, $\theta_{ij}$, and ring angle, $\psi_{ij}$, defined on each edge as in equations~\eqref{eq_spring_stiffness}$-$\eqref{eq_learning_rule}
\begin{subequations} \label{eq_learning_network}
\begin{align} 
    \ddot{\eta}_{ij} &= \left(-\ddot{\mathbf{u}}_i + \ddot{\mathbf{u}}_j \right) \cdot \hat{\mathbf{b}}_{ij}\\
    \ddot{\theta}_{ij} &=  - \omega_p^2 \theta_{ij} - b_p \dot{\theta}_{ij} +\frac{1}{2\ell_p}\left(-\ddot{\mathbf{u}}_i - \ddot{\mathbf{u}}_j \right) \cdot \hat{\mathbf{b}}_{ij} \\
    \dot{\psi}_{ij} &= \gamma \,\text{Relu}\left(\dot{\theta}_{ij} - \tau_0 \right) \, \Theta \left( \dot{\eta}_{ij} - \tau_1 \right) \\
    k_{ij} &= k^0 + \frac{k_{\max} - k_{\min}}{2} \cos \left( 4\psi_{ij} + \alpha_0 \right)
\end{align}
\end{subequations}
Equations \eqref{eq_elastic_network_force}$-$\eqref{eq_learning_network} describe an ADS network, and represent the network version of equations~\eqref{eq_friction}$-$\eqref{eq_learning_rule}. In particular, the edge vectors $\hat{\boldsymbol{b}}_{ij}$ encode the spatial layout and angles between springs, so the coupling between ADSs depends explicitly on the circuit geometry. In Fig.~\ref{fig4}e, we simulate equations~\eqref{eq_elastic_network_force}–\eqref{eq_learning_network} with $\tau_0,\tau_1$ set to $0$ (equation~\ref{eq_learning_network}) and use as initial conditions the experimentally measured node positions and spring orientations for the circuits in Fig.~\ref{fig4}a–d. With these inputs, the learning dynamics of our ADS network experiments are well described by the model without any fit parameters (Fig.~\ref{fig4}e, SI).

Qualitative features of the learning dynamics can be inferred from graph structure alone. For example, the adjoint circuit graph can explain how head and tail interactions determine the relative update rate of neighboring ADSs (Fig.~\ref{fig4}e). The adjoint graph (Fig.~\ref{fig4}e, bottom row) represents the ADSs as nodes, which are joined by an edge if the corresponding ADSs are connected. When parameters are chosen to place the system away from resonance effects, the update rate of an ADS depends primarily on the amplitude difference between the tail and head oscillations. Neighboring springs which apply forces to the tail end of a given ADS can thus be thought of as promoting spring stiffness update, while those connected to the head end inhibit update (Fig.~\ref{fig4}e). This intuitive picture helps explain ADS update dynamics in simple circuits (Fig.~\ref{fig4}e). By composing local promoter-inhibitor motifs, we can construct larger ADS networks with distinct adaptive stiffness properties.  Below, we theoretically and numerically investigate the learning dynamics of mechanical circuits.

\subsection{Pattern finding in ADS circuits}

\begin{figure}[t]
	\centering
    \includegraphics[width=\columnwidth]{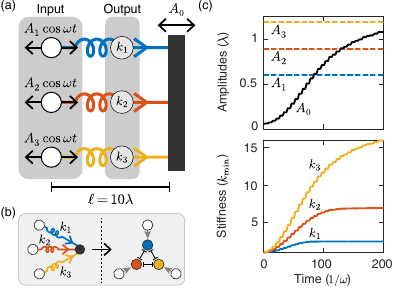}
	\caption{\textbf{3-to-1 ADS circuit dynamics with constant amplitude inputs.}
    (a) Circuit schematic, where each spring receives single-frequency input oscillations with constant amplitudes $A_i$. The spring rest length is $\ell=10\lambda$, where $\lambda$ is the characteristic input displacement scale.
    (b) The circuit graph of the mechanical circuit (left) and its adjoint graph (right) illustrate that each ADS spring inhibits the others (flat arrowheads).
    (c) Numerical integration of equation~(\ref{eq_Nto1}) with $ k_{\min}/(M\omega^2) = 0.02$, $k_{\max}/k_{\min}=20$ and $\gamma/(2 \ell_p) = 60$. Top row: Constant inputs $A_{i}$ and the resulting output amplitude $A_0(t)$. 
    Bottom row: Spring stiffness evolution $k_{i}(t)$; each spring stiffens while $A_j>A_0(t)$ and remains constant for $A_0(t)>A_j$.}
 
	\label{fig4_5}
\end{figure}

By choosing combinations of ADSs which regulate each other appropriately, small ADS networks can be made to identify patterns in stochastic input forces, in a similar manner to simple neural networks~\cite{oja1982simplified}. To see this, we first consider an $N$-to-1 mechanical circuit of ADSs with all displacements constrained to be in the $x$-direction, represented in Fig.~\ref{fig4_5}(a,b) for $N=3$. The $N$ input nodes and ADSs can be labeled by a single index $j=1,...,N$. The input nodes have displacements $x_j(t)$ and the output node has displacement $u(t)$. The additional ADS degrees of freedom are as in equation~\eqref{eq_learning_network}: the pendulum angle $\theta_j(t)$, ring angle $\psi_j(t)$ and spring stiffness $k_j = k_j(\psi_j(t))$. To simplify the analysis, the pendulum parameters are taken to be such that equation~\eqref{eq_oscillator} holds, and specifically, $2\ell_p\dot{\theta}_j = -\dot{u} - \dot{x}_j$. We further set damping and friction coefficients to zero, $b=0,\; \mu=0$, along with the thresholds in the learning rule, $\tau_0=0,\; \tau_1 =0$. Finally, the input nodes are chosen to have prescribed displacements $x_j(t) = A_j \cos\omega t$ with $A_j \geq 0$. Under these conditions, the ADS network equations \eqref{eq_elastic_network_force}$-$\eqref{eq_learning_network} become  
\begin{align}
\begin{split}
    M\ddot{u} &=  \sum_j k_j(x_j - u)\\
    \dot{k}_j &= \frac{\gamma}{2\ell_p} k'(\psi_j) \, \text{Relu} (-\dot{u}-\dot{x}_j)\Theta(\dot{u} - \dot{x}_j)
\end{split}
\label{eq_Nto1}
\end{align}
where $k'(\psi)$ is the derivative of the stiffness function $k(\psi)$ (equation~\ref{eq_spring_stiffness}).

Further progress on equation~\eqref{eq_Nto1} can be made under the assumption that there is a separation of timescales, such that the stiffnesses vary slowly compared to the drive, $|\dot{k}_j|/k_j \ll \omega$ (in our experiments $|\dot{k}|/(k\omega)\sim 0.03$). We therefore first solve for the instantaneous forced response $u(t)$ at fixed $\bs{k}$. Let $\bs{k}, \bs{A}$ be the vectors with components $k_j$ and $A_j$, and let $|\bs{k}|_1 = \sum_j k_j$. The natural frequency of $u(t)$ can then be written as $\omega_s^2 = {|\bs{k}|_1}/{M}$. Defining the dimensionless spring stiffnesses, $\tilde{\bs{k}} = \bs{k} / M\omega^2$, $\tilde{k}_j = k_j / M\omega^2$, the general solution for $u$ is
\begin{align}
	u(t) &= \alpha_1 \cos \omega_s t + \alpha_2 \sin \omega_s t - A_0(\tilde{\bs{k}}) \cos \omega t \\
	A_0(\tilde{\bs{k}}) &= \frac{\bs{k} \cdot \bs{A}}{M\omega^2 -|\bs{k}|_1 } = \frac{\tilde{\bs{k}}\cdot \bs{A}}{ 1 - |\tilde{\bs{k}}|_1}
    \label{eq_A0}
\end{align}
Each spring stiffness is bounded by $k_j \le k_{\max}$ (as in equation~\ref{eq_spring_stiffness}), so
$|\bs{k}|_1 \le N k_{\max}$ and the natural frequency of the
output node satisfies $\omega_s^2 = {|\bs{k}|_1}/{M} \le {N k_{\max}}/{M}$.
To understand the behavior of $A_0$, we assume that the driving frequency $\omega$ is large compared to the natural frequency $\omega_s$, which implies
\begin{align*}
    \omega^2 > Nk_{\max} / M 
\end{align*}
Since $Nk_{\max}/M$ is the maximum possible value of $\omega_s^2$, this condition is equivalently $1 > N\tilde{k}_{\max}$ where $\tilde{k}_{\max} = k_{\max}/M\omega^2$. Since
$|\tilde{\bs{k}}|_1 \le N\tilde{k}_{\max}$ it follows that
$1 - |\tilde{\bs{k}}|_1 \ge 1 - N\tilde{k}_{\max} > 0$. Hence
the denominator of $A_0$ in equation \eqref{eq_A0} is strictly positive so $A_0\geq 0$. Additionally, for $\omega>\omega_s$ and initial conditions such that $\alpha_{i}/A_0=O(1)$ (e.g. starting at rest, 
$u(0)=\dot u(0)=0$), we have
\begin{equation*}
    \dot u(t)\approx A_0\omega\sin(\omega t)\left[1+ O\left(\frac{\omega_s}{\omega}\right)\right]
\end{equation*}
To leading order, the spring updates are then governed by the $k_j$ equations
\begin{align}
\dot{k}_j
  &= \frac{\gamma}{2\ell_p}k'(\psi_j)\,
     \text{Relu}\big[(-A_0 + A_j)\,\omega\sin\omega t \big] \nonumber\\
  &\qquad\qquad\times
     \Theta\big[(A_0 + A_j)\,\omega\sin\omega t \big] \nonumber\\
  &= \frac{\gamma \omega}{2\ell_p} k'(\psi_j)\,
     \text{Relu}(A_j - A_0)\,\text{Relu}(\sin\omega t) 
\end{align}
where the second equality follows from $A_0, A_j \geq 0$. The stiffness $k_j$ therefore updates when $A_j>A_0$ and is constant otherwise.

In certain regimes, these equations exhibit a promoter-inhibitor structure, where the spring stiffnesses regulate each other. To observe this, we focus on a sector in which the ADS stiffnesses are all non-decreasing, $\dot{k}_j\geq 0$. This occurs when $\psi_j$ is such that $k'_j(\psi_j) \geq 0$. Consider a simple example of a $3$-to-$1$ network with constant input amplitudes $0< A_1 < A_2 < A_3$ (Fig.~\ref{fig4_5}). If $A_0 < A_1 < A_2$ initially, both $k_1$ and $k_2$ update and increase. $A_0$ is a monotonically increasing function of the $k_i$, since
\begin{align*}
	\frac{\partial A_0}{\partial \tilde{k}_i} = \frac{A_i}{1- |\tilde{\bs{k}}|_1}  + \frac{\tilde{\bs{k}}\cdot \bs{A}}{(1 - |\tilde{\bs{k}}|_1)^2} > 0 
\end{align*}
so $A_0$ grows as $k_1$ and $k_2$ stiffen. $A_0$ eventually increases until the system reaches a regime
$A_1 \leq A_0 < A_2$, at which point spring~1 is inhibited
($A_1 \leq A_0$) while spring~2 continues to promote stiffness growth
($A_2 > A_0$).
Thus high-amplitude inputs act as promoters of stiffness growth,
whereas lower-amplitude inputs become inhibited once they are exceeded by the output
amplitude $A_0$. This reproduces the
promoter-inhibitor dynamics observed by numerically simulating equation \eqref{eq_Nto1} (Fig.~\ref{fig4_5}c). 

\begin{figure}[h]
	\centering
    \includegraphics[width=\columnwidth]{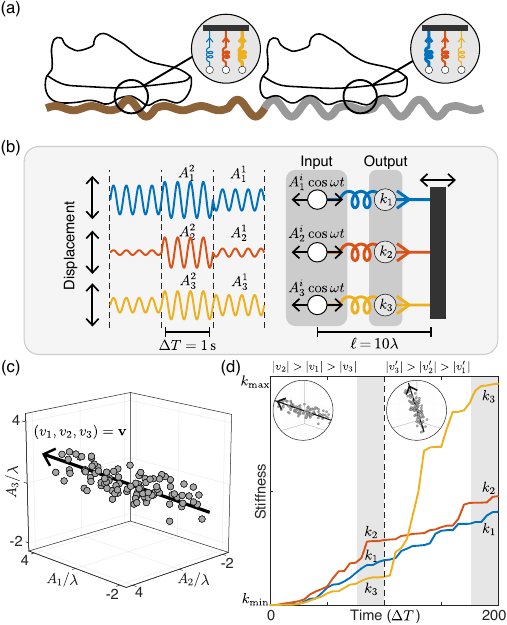}
	\caption{\textbf{Autonomous pattern finding with simulated} mechanical circuits. 
    (a)~Pattern finding in a simplified toy model of a shoe walking on a surface. 
    (b)~The network receives single-frequency oscillations with a sequence of amplitudes, $\bs{A}^i = (A^i_1, A^i_2, A^i_3)$, in time intervals $\Delta T$, representing environmental interactions ($\ell=10\lambda$ is the spring rest length). 
    (c)~The amplitudes, $\bs A^i$, are sampled i.i.d. from $\bs A^i/\lambda\sim\mathcal N(\mathbf 1,\Sigma)$, where  $\Sigma=2\,\mathbf{v}\mathbf{v}^{\mathsf T}+\tfrac12\!\left(I-\mathbf{v}\mathbf{v}^{\mathsf T}\right)$ and $\mathbf{v}\sim\mathcal U(S^2)$ represents the underlying pattern in the environment. $A^i_j<0$ corresponds to an oscillation phase-shifted by $\pi$.
    (d)~The spring stiffnesses are initialized at $k_{\min}$ and evolve according to equations~\eqref{eq_elastic_network} and \eqref{eq_learning_network} (see SI for parameters). The ordering of $(k_1,k_2,k_3)$ updates to match the ordering of the components $(|v_1|, |v_2|, |v_3|)$ with probability~83$\%$. When the pattern changes, $\mathbf{v} \rightarrow \mathbf{v}'$ (dashed line at $t=100\,\Delta T$), the stiffnesses adapt and learn the new ordering with probability~62$\%$. Success probabilities are obtained from 200 repetitions.
    } 
	\label{fig5}
\end{figure}

The self-regulation properties of the many-to-one ADS network indicate that these mechanical circuits could be used for learning. We consider a self-learning task that can be viewed as a toy model of a shoe walking on a textured surface (Fig.~\ref{fig5}a). As before, the stochastic environment is realized by driving a 3-to-1 circuit using a sequence of displacement waves of fixed frequency $\omega$, in the horizontal direction. The wave amplitudes are now taken to be time-varying and random, with underlying correlations (Fig.~\ref{fig5}b). Specifically, during each interval $i$
(of duration $\Delta T$), input node $j$ has prescribed displacement
$x_j(t)=A^i_j\cos(\omega t)$, where the amplitudes
$\bs{A}^i=(A^i_1,A^i_2,A^i_3)$ are sampled i.i.d. from a multivariate
Gaussian distribution,
$\bs{A}^i/\lambda \sim \mathcal{N}(\mathbf{1},\Sigma)$ and $\ell=10\lambda$ is the rest-length of each ADS.
The covariance matrix is chosen to have largest variance in a direction $\mathbf{v}=(v_1, v_2, v_3)$, where $\mathbf{v}$ is sampled uniformly at random on the unit sphere $\mathbf{v}\sim \mathcal{U}(S^2)$. More precisely, $\Sigma$ has the form $\Sigma=2\,\mathbf{v}\mathbf{v}^{\mathsf T}+\tfrac12\!\left(I-\mathbf{v}\mathbf{v}^{\mathsf T}\right)$, so $\mathbf{v}$ is the top eigenvector of $\Sigma$. The resulting distribution of amplitude vectors, $\bs{A}^i$, is therefore skewed in the direction of $\mathbf{v}$ (Fig.~\ref{fig5}c). The amplitudes $\bs{A}^i$ drawn from this distribution can sometimes be negative; these values correspond to oscillations phase shifted by $\pi$, since $-A\cos\omega t = A\cos(\omega t+\pi)$. 

The stiffness pattern within a simulated network evolves to reflect the statistical properties of the inputs $\bs{A}^1,\bs{A}^2,...$ (Fig.~\ref{fig5}d). In particular, initializing each spring at $k_i = k_{\min}$, the ordering of the ADS stiffnesses $k_1,k_2, k_3$  updates to match the ordering of the components of the top eigenvector, $|v_1|,|v_2|,|v_3|$, (Fig.~\ref{fig5}c) with probability $83\%$ (2 s.f., 200 trials). A change in environmental conditions can be captured by sampling a new $\mathbf{v}' \sim \mathcal{U}(S^2)$ and setting the covariance matrix to be $\Sigma=2\,\mathbf{v}'\mathbf{v}'^{\mathsf T}+\tfrac12\!\left(I-\mathbf{v}'\mathbf{v}'^{\mathsf T}\right)$. Adapting to this new environment is more difficult since the network must first forget its previous state. In simulations, the network correctly learns the ordering of the components $(|v'_1|, |v'_2|, |v'_3|)$ (Fig.~\ref{fig5}d) with probability $62\%$ (2 s.f., 200 trials). For comparison, without memory, the only information about $\mathbf{v}$ comes from the amplitudes $\bs{A}/\lambda$, chosen from $\mathcal{N}(\mathbf{1}, \Sigma)$. The probability that the ordering of $|A_1|, |A _2|, |A_3|$ matches the ordering $|v_1|,|v_2|,|v_3|$ can be calculated numerically, and is $24\%$ (2 s.f.). 

This computational task represents an idealized scenario, in which the environmental inputs consist of waves with equal frequency (Fig.~\ref{fig5}b) and the spring stiffnesses are initialized at the minimum value $k_{\min}$ (Fig.~\ref{fig5}d). Despite these simplifications, this example demonstrates the ability of mechanical circuits to compute features of a changing environment. Furthermore, the learning dynamics is driven by the input signals alone. Because springs in the simulated circuit tend to stiffen where $|v_i|$ is high and remain compliant where it is low, the learned state has a particular mechanical signature which could be harnessed for function.

\subsection{Large mechanical circuits}

\begin{figure*}[t]
	\centering
	\includegraphics[width=\textwidth]{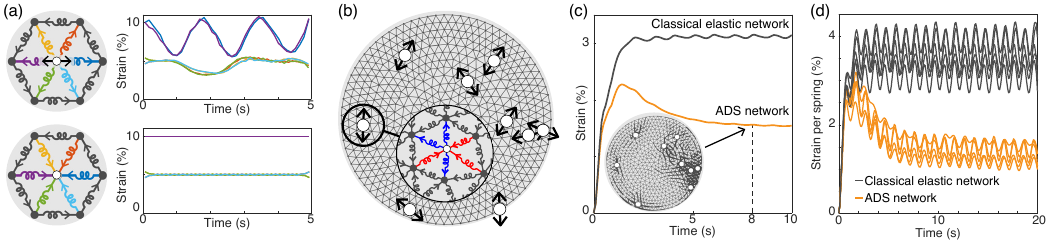}
	\caption{\textbf{Adaptive damping in simulated mechanical circuits.} (a)~Directional adaptivity controls how strain is transmitted through an ADS network. ADSs adjacent to oscillating nodes with large out-degree (top row, white node) update and redistribute strain, thus reducing their average strain over oscillation cycles (colored curves). ADSs around forced nodes with large in-degree (bottom row) experience higher average strain. In the latter case, the strain is effectively caged (bottom row). In both cases, the ADSs are initialized in their softest state ($k_i = k_{\min}$ for all ADSs), and the input oscillation is horizontal (top row, black arrow).
    (b,c)~Strain redistribution within a large ADS network, initially in its softest state. (b)~Caging is unlikely in large ADS networks with randomly chosen forcing nodes (white) and random ADS orientations. Forcing nodes are typically adjacent to both inward pointing ADSs (inset, red) and outward ADSs (inset, blue), which enable strain redistribution. (c)~The ADS network experiences a lower average strain (orange curve) than a non-adaptive elastic network (black curve) undergoing the same input oscillations. Stiffening ADSs redistribute strain (inset, thicker edges denote stiffer springs). (d)~A classical elastic network and an ADS network are initialized with identical spring stiffnesses and subject to the same, randomly chosen forcing protocol. Over several such trials, the ADS network distributes strain and thus experiences lower overall strains than a classical non-adaptive network. Simulations correspond to equations~\eqref{eq_elastic_network_force}–\eqref{eq_learning_network} (see SI for parameters).} 
	\label{fig5_pre}
\end{figure*}

Large ADS networks can be thought of as ensembles of smaller mechanical circuits. In particular, a single node together with its neighboring ADSs is an elementary mechanical circuit, from which larger networks can be constructed (Fig.~\ref{fig5_pre}a). A key property of nodes within such elementary circuits is the number of adjacent ADSs which point inwards or outwards (Fig.~\ref{fig5_pre}a). The oscillation of nodes with large out-degree causes adjacent ADSs to update, ultimately leading to the redistribution of strain throughout the network (Fig.~\ref{fig5_pre}a, top). On the other hand, friction and directionality combine to cage oscillations of nodes with zero out-degree (Fig.~\ref{fig5_pre}a, bottom). In a large network with randomly oriented ADSs, forcing nodes are unlikely to have zero out-degree (Fig.~\ref{fig5_pre}b,c). When forced at random locations, local spring update will therefore typically cause such ADS networks to distribute loads throughout the system.

The adaptive damping behavior of a large ADS network contrasts with the dynamics of a classical non-adaptive elastic network with the same connectivity (Fig.~\ref{fig5_pre}c,d). To compare these cases, we initialize each network so that all springs have stiffness $k_{\min}$ at time $0$. A set of nodes, $i_1,i_2,\ldots,i_m \in V$ within the network are chosen uniformly at random to experience an oscillating force. For each $s=1,\ldots,m$, node $i_s$ experiences a force $F\cos \omega t \, (\cos \phi_s, \sin\phi_s)$. The magnitude $F$, and frequency $\omega$, are identical for each node, but the direction $\phi_s$ of the force is node-dependent, and is chosen uniformly at random. The stiffening of springs in an ADS network distributes forces leading to a lower average strain than a non-adaptive network. This effect occurs over several trials (Fig.~\ref{fig5_pre}d), thus confirming the adaptive damping behavior of large ADS networks.

\section{Discussion}

ADS directionality gives rise to an exponentially large design space of mechanical circuits. 
The size of this space suggests that networks could be chosen to solve a range of learning tasks. The small network motifs illustrated here (Figs.~\ref{fig4}-\ref{fig5_pre}) exhibit learning dynamics, and can serve as building blocks of more general networks. The design space can be broadened further by utilizing additional degrees of freedom within the system, such as allowing non-constant node masses. However, understanding the mapping between network structure and learning task is a key challenge of the mechanical circuit approach. Machine learning methods, which have been successfully used for inverse design of materials~\cite{ha2023rapid}, represent a potential solution for this network optimization problem.

To conclude, using adaptive directed springs we construct battery-free mechanical circuits which can learn information about their environment. This learning behavior is embodied in the mechanical domain itself, allowing for environmental interaction unmediated by a separate sensory layer~\cite{yang2022memristor,prabhakar2022mechanical}. Our circuits thus represent a mechanical approach to the problem of designing materials with embodied intelligence~\cite{el2022mechanical,hu2018small, morton2023autonomous,hauser2011towards}. Although our physical realization of an ADS involves several components, the individual mechanisms of a spring, pendulum and ratchet, are generic, and can be further augmented with a variety of nonlinear mechano-dynamical systems~\cite{berry2022controlling,coulais2017static,sounas2017non,fruchart2021non,brandenbourger2021limit, brandenbourger2019non}. The genericity of the underlying components suggests that ADS dynamics could be reproduced in different domains and length scales~\cite{robin2023long,hughes2019wave, preston2019digital, schaffter2022standardized}, such as fluidic~\cite{robin2023long, woodhouse2017active} and electromechanical~\cite{el2022mechanical} systems. A particularly interesting question is whether ADS-type behavior plays a role in adaptivity in biological contexts~\cite{rashid2023mechanomemory, foster2019cytoskeletal}. The framework developed here could thus improve our understanding of self-learning in living matter. More generally, our results provide a set of mechanical tools for constructing self-learning materials which continually adapt and self-optimize.

\section{Acknowledgements}
V.P.P., I.H. and M.P. conceptualized this work. V.P.P. performed simulations and analytic calculations. I.H. designed and performed experiments. All authors contributed to interpreting data, and writing the manuscript. This work was financially supported by a Stanford Science Fellowship (V.P.P.), a Tau Beta Pi Fellowship (I.H.), a HHMI Faculty Fellowship (M.P.), Bio-Hub Investigator Fellowship (M.P.), Schmidt Innovation Fellowship (M.P.), Moore Foundation Research Grant (M.P.) and ARIA Nature Computes Better Program (M.P.). 
 
\bibliography{references}

\end{document}